\documentclass[%
 reprint,
 amsmath,amssymb,
 aps,
]{revtex4-1}

\usepackage{graphicx}
\usepackage{dcolumn}
\usepackage{bm}

\begin{document}

\title{PAW mediated \textit{ab initio} simulations on linear response phonon dynamics of anisotropic black phosphorous monolayer for thermoelectric applications}
\author{Sushant Kumar Behera and Pritam Deb}
 \email{Corresponding Author: pdeb@tezu.ernet.in}
 \affiliation{Advanced Functional Material Laboratory, Department of Physics, Tezpur University (Central University), Tezpur-784028, India.}

\date{\today}

\begin{abstract}
The first order standard perturbation theory combined with \textit{ab initio} projector augmented wave operator challenges the realization of the standard Sternheimer equation with linear computational efficiency. This efficiency motivates us 
to describe the electron-phonon interaction in two-dimensional (2D) black phosphorous monolayer using generalized density functional perturbation theory (DFPT) with Boltzmann transport theory (BTE). Subsequently, linear response 
phonon dynamic behaviour in terms of conductivities, seebeck coefficients and transport properties are focused for its thermoelectric application. The analysis reveals the crystal orientation dependence via structural anisotropy and 
the density of states of the monolayer structure. Momentum dependent phonon population dynamics along with the phonon linewidth are efficient in terms of reciprocal space electronic states. The optimized values of thermal 
conductivities of electrons and Seebeck coefficients act as driving force to modulate thermoelectric effects. Figure of merit is calculated to be $\sim$ 0.074 at 300 K and $\sim$ 0.152 at 500 K of the MLBP system as a function of the power 
factor. The value of lattice thermal conductivity is 37.15 W/mK at room temperature and follows the inverse dependency with temperature. With the anticipated superior performance, profound thermoelectric applications can be 
achieved particularly in the monolayer black phosphorous system.
  
\begin{description}
\item[keywords]
black phosphorous monolayer, linear response dynamic behavior, phonon population
\end{description}
\end{abstract}

\maketitle

\section{Introduction}
Quantum confinement effect plays primary role in low dimensional semiconductors to perform efficiently as thermoelectric materials \cite{1}. The carrier energy tunes rapidly the electronic states of such reduced dimensional systems. 
As a result, Seebeck coefficient is automatically enhanced for better performance \cite{2}. Figure of merit (ZT), a dimensionless factor, quantiﬁes thermoelectric device efﬁciency relating the Seebeck coefﬁcient (i.e. the thermopower) 
to electronic thermal conductivity. The lesser thermal conductivity value along with relatively higher thermopower and electrical conductivity values are robust aspects for high efﬁciency thermoelectric materials. The 
efﬁciency improvement is mainly controlled due to the sharp peaked electronic density of states (DOS). Nanotechnology has been applied extensively to improve the thermoelectric performance since the past two decades \cite{3,4}. 
Few of the nanostructures \cite{5,6,7,8}, 2-dimensional electron gas (2 DEG) \cite{9,10} and nanowires \cite{11} have all been reported for superior thermoelectric and transport properties. However, it is difficult task to control 
the dimensional scaling of such structures to achieve superior performance cost effectively. In the line of search for effective structures as enriched thermoelectric and phonon transport performance, natural two dimensional (2D) 
materials with finite bandgap (i.e. semiconductors or semimetals), low energy dispersion, high carrier mobility and minimized phonon modes are considered as suitable candidates \cite{12,13,14,15,16,17,18}.   \\

\begin{figure}[th!]     
\centering           
\includegraphics[width=8cm,height=6cm]{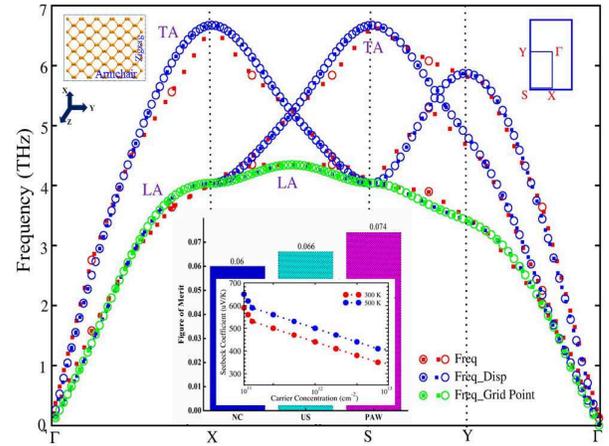}
\caption{\label{Fig:wide}The schematic illustration of PAW effect on thermal conductivity value of MLBP for thermoelectric application.}
\end{figure} 

Recently, monolayer black phosphorous (MLBP) known as an allotrope of bulk black phosphorous, a 2D material family, has appeared in this line of research \cite{19,20,21,22,23}. MLBP possesses puckered honeycomb lattice of phosphorous 
atoms with low symmetry and highly anisotropy resulting many interesting and applied active benefits of the structure \cite{24,25,26,27}. The Seebeck coefficient and phonon modes are directly dependent on this anisotropic 
electronic structure resulting better thermoelectric and phonon transport performance. In this aspect, experimental findings on multi-layer or monolayer black phosphorous have been reported to realize the theoretical 
predictions \cite{28,29,30,31}. More recently, electron-phonon interaction in phosphorene has been performed via first-principle calculations based on DFPT and Wannier interpolation with norm conserving pseudo potential \cite{32}.   \\

\begin{figure}[th!]     
\centering           
\includegraphics[width=8cm,height=6cm]{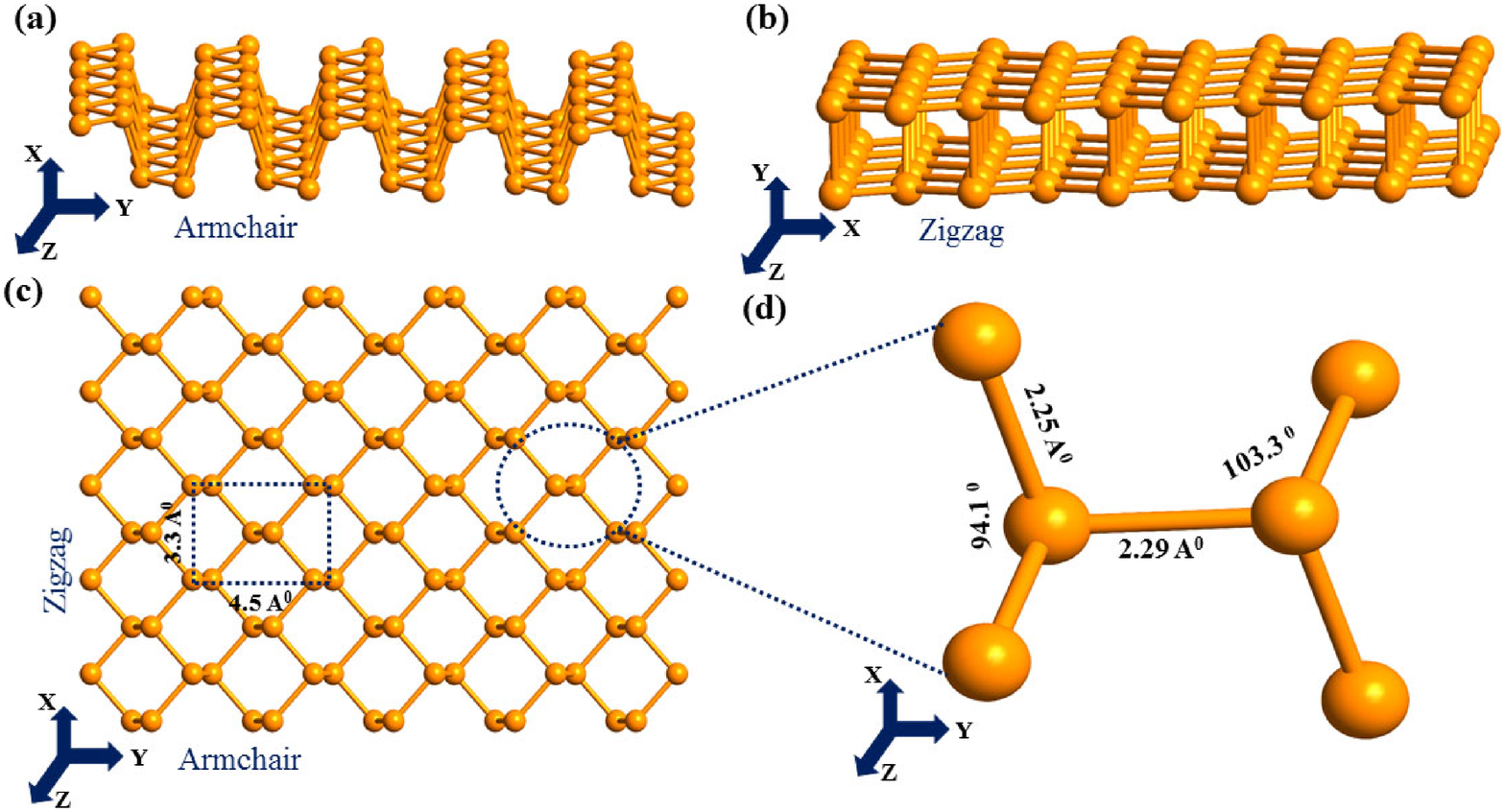}
\caption{\label{Fig:wide}The optimized geometry of phosphorene monolayer in (a) armchair and (b) zigzag direction shown as side view. The surface aerial view of the monolayer sheet 
from both the direction shown in (c) and the angle and bond length among the phosphorous atom in (d).}
\end{figure} 

In the line of understanding, first order Kohn-Sham equations written in the form of a perturbation series are used to realize the basic physics behind standard perturbation theory of Sternheimer equation \cite{33} in the 
perspective of ab initio DFPT method through self-consistency field. Besides, first order Kohn-Sham Hamiltonian is linearly dependent on electron wave function in strongly correlated electron systems like 2D material sheet 
resulting manifold coupling between conduction and valence bands \cite{34}. Interestingly, projector augmented wave operator is the only option to express such linear dependency in case of the perturbation stage to validate the 
Sternheimer equations. Interestingly, this first order response of the Kohn Sham Hamiltonian will support to implement the projector augmented wave pseudopotential based DFPT algorithm to calculate thermal and phonon responses 
with linear computational efficacy in MLBP system. Theoretically, we are still lacking to implement the linear response phonon dynamics of in 2D monolayer sheets of anisotropic materials in the framework of PAW based DFPT 
technique. Thus, stimulating research endeavor can be implemented to evaluate the phonon mediated dynamic behaviour and potential thermoelectric performance of MLBP.    \\

\begin{figure}
\centering           
\includegraphics[width=8cm,height=7cm]{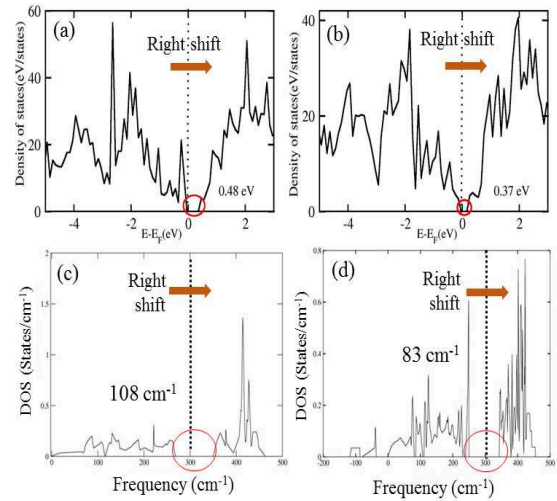}
\caption{\label{Fig:wide}The distribution of electronic density of states (DOS) of the monolayer at (a) 300 K and (b) 500 K. The dotted black line represents the Fermi level and the red dotted circle shows the 
closing of gap due to Increasement in temperature. The distribution due to phonon states shown at (c) 300 K and (d) 500 K of the same optimized structure. The dotted black line presents the frequency line where states are zero.}
\end{figure} 

In this manuscript, we use \textit{ab initio} DFT with projector augmented wave (PAW) pseudopotential method supported by Perdew-Burke-Ernzerhof generalized gradient approximation (PBE-GGA) functional to predict electronic and linear 
response phonon transport behaviour, which self-consistently takes into account for such anisotropic materials. In the context of anisotropic semiconductors, it is interesting to know the feasibility of PAW pseudopotential based 
DFPT method to analyse the linear phonon dynamics, which presents the characteristic coupling between conduction and valence bands. We address specifically such problem in this current simulation work. Here, BTE is implemented for 
thermoelectric properties taking the Boltztrap code. The Seebeck coefficient, electronic thermal conductivities, carrier mobility and power factor are considered during the calculation. Moreover, the linear response phonon 
interaction and dynamic behaviour are calculated for MLBP including the phonon population density with respect to the reciprocal k space symmetry points. Our results, obtained from projector operators and generalized gradient 
functional in the monolayer sheet, are consistent and superior than the previously reported electron-phonon interaction \cite{32} and thermal transport \cite{31} data showing that the momentum-resolved phonon mediated linear response 
behaviour of MLBP through self-consistent ab initio DFPT calculations.  \\ 

\section{Methodology}
The electronic structure of ML black phosphorous is studied using DFT calculation using Quantum Espresso codes \cite{35} along with PAW pseudopotential \cite{36} and the PBE functional within the generalized gradient approximation 
(GGA) \cite{37}. The van der Waals (vdW) interaction has been considered for the monolayer structure \cite{38,39}. A Monkhorst mesh of $9\times9\times1$ k-points is used for geometry optimization with 540 eV as plane wave cutoff energy. 
Optimization iteration process is followed until the total force converged to $\preceq$ 0.001 eV${\AA{}}^{-1}$. Supercells with lattices of 12 $\AA{}$ in the z-direction is considered to neglect the periodic interaction among 
the surface images of the monolayer sheet structures. We use $27\times27\times1$ k-mesh for electronic structure calculations. The phonon mode related simulations are performed within the framework of density functional perturbation 
theory \cite{40} with $27\times27\times1$ k point mesh. The transport phenomena has been studied using Boltzmann transport equation (BTE) \cite{41}. 

\section{Results and Discussion}
The optimized monolayer geometries are determined using quasi Newtonian algorithm. The structures of the layers are shown in Fig. 2. To understand the origin and control of electron states and phononic states at active sites in 
monolayer and their distribution, the total densities of states (TDOSs) and phonon density of states are performed. Overlapping states are observed from the plots of density of states (Fig. 3) showing the active behavior near 
the Fermi region and the gap near to the active sites. Presence of localized electrons on the P edge of monolayer sheet has contributed to the overlapping states at Fermi level within conduction band with confinement and 
delocalization of the phosphorous (P) atoms, resulting the dynamic behaviour of the sites. The schematic of the work has been reflected in Fig. 1. \\

\begin{figure}
\centering           
\includegraphics[width=8cm,height=7cm]{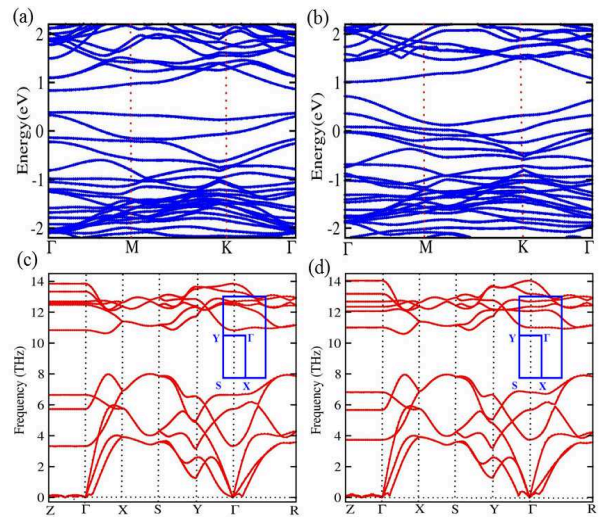}
\caption{\label{Fig:wide}Electronic band structure of (a) 300 K and (b) 500 K of the monolayer phosphorene. The high symmetry points are Γ, M, K and Γ in reciprocal space. Phononic band structure of the 
same monolayer at (c) 300 K and (d) 500 K. The high symmetric frequency points are taken Z, $\Gamma$, X, S, Y, $\Gamma$ and R in reciprocal space. }
\end{figure} 
The electronic (DOS) (shown in Fig. 3 (a) and (b)) with step-like features are observed near the Fermi level and a slight right shifting upon increasing temperature to 500 K because of highly anisotropic behaviour. We observe 
similar horizontal level in optimum band edges of both conduction and valence band laterally zigzag direction indicating possibility to improve the value of Seebeck coefficient. This potential finding is worthy enough for the 
2D monolayer as thermoelectric material, unlike its bulk counterpart (i.e. black phosphorous).   \\

Optical phonon contribution is significantly low in the phonon density of states (Fig. 2 (c) and (d)). We notice slight peak shifting near the band edges from 108 to 83 c$m^{-1}$, which is occurred due to anisotropic band structure 
of MLBP. To correlate the bands due to electronic states distribution and phonon states with DOS pattern, we have plotted the band diagram in both cases of electronic band and phononic bands (shown in Fig. 4). The band gap is 
corroborated with its states calculated from DOS pattern.   \\

\begin{figure}
\centering           
\includegraphics[width=8cm,height=6cm]{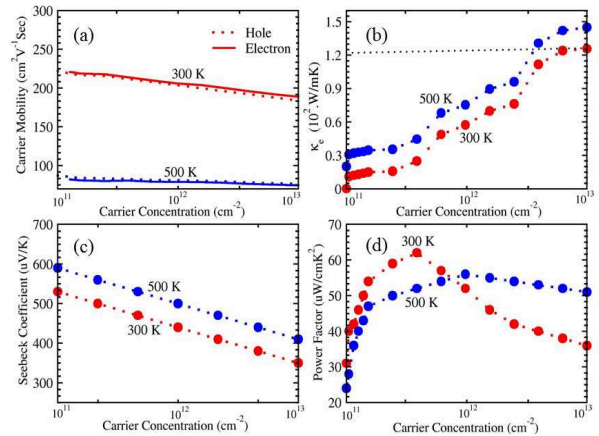}
\caption{\label{Fig:wide} The calculated (a) carrier mobility, (b) electronic thermal conductivity, (c) Seebeck coefficient and (d) thermoelectric power factor along the zigzag direction of MLBP at two temperatures.}
\end{figure} 

The electrons and hole mobilities are plotted as a function of carrier concentration from the shifting of the Fermi level at room temperature (Fig. 5(a)). The phonon mediated carrier mobility of MLBP is $\sim$ 212 c$m^2$/Vs at 300 K 
corroborating with experimentally verified results of few layers of BP \cite{42}. In Fig. 4(b) we present the calculated electronic thermal conductivity of the MLBP with a higher chance of predictability at higher temperature.      \\

In Fig. 5 ((c) and (d)), the Seebeck coefficient and power factor have been plotted as a function of carrier concentrations at 300 and 500 K. The Seebeck coefficient is dominant and the thermoelectric power factor achieves 
~60 μW/cm-$K^2$ at room temperature. The power factor values at 300 K and 500 K have been taken to calculate the figure of merit of the material as thermoelectric application. The thermoelectric materials are defined by the 
figure of merit (ZT), given as ZT=$S^2{\sigma}k^{-1}{\Delta}$T, depending on seebeck coefficient (S), electrical conductivity ($\sigma$), electronic thermal conductivity (k) and the temperature gradient ($\Delta$T). The formula for 
ZT= (PF).$k^{-1}{\Delta}$T is more simplified by considering ($S^2{\sigma}$) as power factor (PF) at the particular temperature gradient to generate electricity \cite{43,44}. The optimal values of ZT is calculated to be approximately 
$\sim$ 0.074 at 300 K and $\sim$ 0.152 at 500 K for the MLBP.  \\

\begin{figure}
\includegraphics[width=9cm,height=4cm]{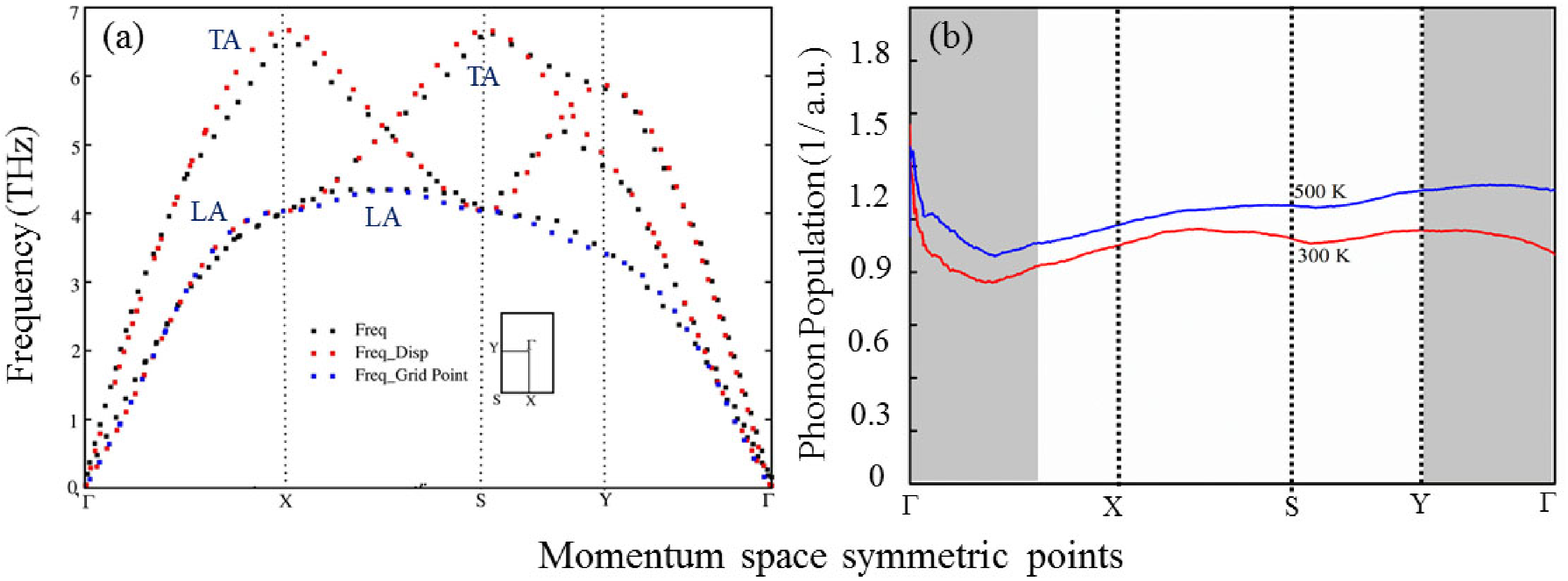}
\caption{\label{Fig:wide}(a) Phonon band structure along the momentum space high-symmetry ($\Gamma$-M-K-$\Gamma$) points. The solid lines are summed over all phonon branches and the dashed line is the sum of all acoustic (i.e. longitudinal 
acoustic (LA) and transverse acoustic (TA)) branches. (b) momentum-resolved first-principles based calculations of phonon population at two different electronic temperatures, the gray areas indicate the q range.}
\end{figure} 

Contributions from the phonon modes in case this 2D system is ignored due to the inversion symmetry. Fig. 6 (a) shows phonon band structure along the high-symmetry and population density of each phonon mode (Fig. 6(b)). 
The phonon density is varying significantly at different symmetric points for all phonon branches of the ML structure at 300 and 500 K along the zigzag direction. \\

We calculate the variation in individual phonon population density as a function of their reciprocal k space and reveal the transport property along the zigzag direction (Fig. 6(b)). The optimized phonon density is determined 
to be 0.95 at room temperature and 1.12 at 500 K near M and K points. Here, the results estimate the relative effectiveness of low dimensional structures in affecting their transport properties. The comparison of the linear 
response phonon population and the phonon band structures (Fig. 6) shows analogous momentum space dependency. Quantitatively, the calculated phonon population is enriched by a factor of 7 on increasing the electronic temperature 
to 500 K. Here, electrons relax to the CBM by consequent phonon scattering. The electronic temperature changes dynamically adjusting the phonon coupling strength with a finite CBM frequency difference at M and K points of more 
than 300 c$m^{-1}$ (Fig. 6(a)). Therefore, the quantitative difference of the phonon population dynamics and the reciprocal space band structure are in good agreement with each other.  \\

\begin{figure}
\centering           
\includegraphics[width=6cm,height=6cm]{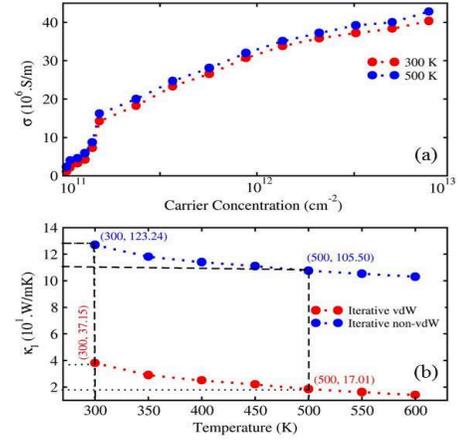}
\caption{\label{Fig:wide}(a) electrical thermal conductivity of electrons and (b) lattice thermal conductivity of phonons of MLBP system along zigzag direction.}
\end{figure} 

The thermoelectric properties of MLBP has been calculated using BTE for electrons at a fixed scattering time. Boltztrap code is used to obtain electrical thermal conductivity (σ) depending on the relaxation-time. In this regard, 
deformation potential theory is implemented to get the relaxation time in case of the MLBP system due to lack of experimental support for the electrical conductivity value. Fig. 7 (a) shows the variation of conductivity values with respect 
to the carrier concentration at 300 and 500 K along zigzag direction. Similarly, the lattice thermal conductivity is obtained from the phonon based BTE via iterative method with van der Waals (vdW) correction and without vdW 
correction. In case of MLBP, strong anisotropic nature tunes its intrinsic values as a function of temperature (shown in Fig. 7 (b)). The thermal conductivity value of MLBP at is found to be 37.15 W.m/K along the zigzag direction at 
room temperature, which is approximately two times greater than the reported value along the armchair direction. As per the theory, temperature rise above room temperature modulates the lattice vibration resulting lattice softening. 
Thus, the atomic arrangement gets disturbed and stiffness decreases, which shrinks the sound velocity and hence the lattice thermal conductivity. \\ 

\section{Conclusion}
In summary, the momentum-resolved phonon mediated linear response behaviour can be understood by examining the phonon scattering of MLBP from first principle calculations taking PAW pseudopotential and PBE-GGA functional. 
Carrier mobility and optical phonon contribution are supporting the band shifting and thermoelectric functionality along zigzag direction due to highly anisotropic nature of the monolayer surface. The phonon scattering rates 
reveals the linear scale directional dependence of the lattice dynamics. The estimated carrier mobility and power factor are found to be 212 c$m^2$/Vs and around ~60 μW/cm-$K^2$ at room temperature, respectively, which are significant 
for intrinsic transport property. Increasing trend of figure of merit and the reduced value of seebeck coefficient supports monolayers to be more favorable than their bulk counterpart, indicating the positiveness of nanostructuring 
MLBP for thermoelectric performance. The results in this study justifies superior performance in thermoelectric applications of monolayer black phosphorous.\\

\begin{acknowledgments}
SKB acknowledges DST, Govt. of India for providing INSPIRE Fellowship. The authors acknowledge Tezpur University for providing HPCC facility.
\end{acknowledgments}

\nocite{*}

\bibliography{manuscript}

\end{document}